# Beyond Pairwise: Nonparametric Kernel Estimators for a Generalized Weitzman Coefficient Across k Distributions

Omar Eidous*  and  Noura Almasri

omarm@yu.edu.jo        nourahussin997@gmail.com

Department of Statistics - Yarmouk University

Irbid - Jordan

## Abstract

This papers presents a generalization of the Weitzman overlapping coefficient, originally defined for two probability density functions, to a setting involving $k$ independent distributions, denoted by $\Delta_k$. To estimate this generalized coefficient, we develop nonparametric methods based on kernel density estimation using $k$ independent random samples ($k \geq 2$. Given the analytical complexity of directly deriving $\Delta_k$ using kernel estimators, a novel estimation strategy is proposed. It reformulates $\Delta_k$ as the expected value of a suitably defined function, which is then estimated via the method of moments and the resulting expressions are combined with kernel density estimators to construct the proposed estimators. This method yields multiple new estimators for the generalized Weitzman coefficient. Their performance is evaluated and compared through extensive Monte Carlo simulations. The results demonstrate that the proposed estimators are both effective and practically applicable, providing flexible tools for measuring overlap among multiple distributions.

* Corresponding Author

## 1. Introduction

Overlapping coefficients are widely used to quantify the similarity or agreement between two or more probability density functions ($pdfs$). Among the most commonly studied measures are the Matusita, Morisita, and Weitzman coefficients, typically denoted by $\rho$, $\lambda$ and $\Delta$ respectively (Eidous and Maqableh, 2024 and Eidous and Ananbeh, 2025). Of these, the Weitzman coefficient $\Delta$ is particularly appealing due to its interpretability and ease of generalization beyond two distributions (Eidous and Alsheyyab, 2025, 2026).

Given two continuous $pdfs$ $f_1(x)$ and $f_2(x)$, the classical Weitzman overlap coefficient is defined as:

$$\Delta_2 = \int min\{f_1(x), f_2(x)\}dx.$$

This integral represents the area under the point-wise minimum of the two densities, and its value lies in the interval [0, 1]. A value of 0 indicates no overlap (complete dissimilarity), whereas a value of 1 implies perfect agreement between the distributions.

This concept naturally extends to more than two distributions. For instance, the overlap among three $f_1(x)$, $f_2(x)$ and $f_3(x)$ is given by:

$$\Delta_3 = \int min\{f_1(x), f_2(x), f_3(x)\}dx.$$

In general, for $k$ $pdfs$ $f_1(x)$, $f_2(x)$, ..., $f_k(x)$, the generalized Weitzman coefficient is:

$$\Delta_k = \int min\{f_1(x), f_2(x), ..., f_k(x)\}dx.$$

Originally, $\Delta_2$ introduced in the context of income distribution (Weitzman, 1970), the $\Delta_2$ coefficient has since been adopted in various fields, including applied statistics (Inman and Bradley, 1989), economics (Milanovic and Yitzhaki, 2002), medicine (Mizuno et al., 2005), and ecology (Núñez-Antonio et al., 2018). Moreover, it plays a role in nonparametric goodness-of-fit testing (Alodat et al., 2022).

Several prior works have explored both parametric and nonparametric estimation of $\Delta_2$. Parametric methods typically assume known functional forms for the distributions (e.g., normal or Weibull) and rely on maximum likelihood estimation (Eidous and Abu Al-hayja's, 2023). For example, Inman and Bradley (1989) considered $\Delta_2$ under equal-variance normal distributions, while others such as Reiser and Faraggi (1999) and Mulekar and Mishra (2000) derived confidence intervals under various assumptions.

However, in many real-world applications, the underlying density functions may be unknown or too complex to specify parametrically. In such cases, nonparametric estimation becomes essential. Kernel density estimators (KDEs) offer a flexible tool for this purpose. Given a random sample $X_{i1}, X_{i2}, \ldots X_{in_i}$ from a continuous $pdf\, f_i(x)$, the kernel estimator of $f_i(x)$ is (Silverman, 1986),

$$\hat{f}_i(x) = \frac{1}{n_i h_i} \sum_{j=1}^{n_i} K\left(\frac{x - X_{ij}}{h_i}\right), \quad -\infty < x < \infty,$$

where $K(,)$ is a kernel function (typically standard normal), and $h_i$ is a bandwidth parameter, often chosen using Silverman's rule of thumb:

$$h_i = (4/3)^{1/5} A_i n_i^{-1/5},$$

where $A_i = min\{S_i,\ IQR_i/1.34\}$. Here, $S_i$ and $IQR_i$ are the sample standard deviation and interquartile range, respectively.

In the two-sample case, nonparametric estimator of $\Delta_2$ is constructed by substituting KDEs into the overlap integral:

$$\widehat{\Delta}_2 = \int_{-\infty}^{\infty} min\{\hat{f}_1(x), \hat{f}_2(x)\} dx.$$

Such methods have been examined by Clemons and Bradley (2000), Schmid and Schmidt (2006), Eidous and Ananbeh (2024) and others using applications ranging from gender income gaps to environmental data.

Recent studies (Eidous and Al-Talafha, 2020; Eidous and Alshorman, 2023; Magableh and Eidous, 2024; and Eidous and Daradkeh, 2024) have proposed new approaches for estimating $\Delta_2$ by expressing it either as an expected value or through numerical approximations such as the trapezoidal and Simpson's rules.

Despite that the generalization study by Eidous and Alsheyyab (2025, 2026) shares a similar goal with the present work, namely analyzing $k$ independent samples, the two approaches differ substantially. While the present study adopts a nonparametric kernel-based method, which does not assume specific functional forms for the underlying distributions, the work of Eidous and Alsheyyab is restricted to the case of normally distributed populations.

Building on this body of works, the current study proposes a fully nonparametric framework for estimating the generalized overlap coefficient $\Delta_k$ across $k$ independent distributions. This generalization is both theoretically significant and practically valuable, as it allows for quantifying similarity among multiple groups, an essential task in comparative studies, cluster analysis, and related applications.

## 2. Illustrative Numerical Example

Let $X$ be a continuous random variable with the Weibull probability density function ($pdf$) given by,

$$f_X(x) = \frac{\beta}{\alpha}\left(\frac{x}{\alpha}\right)^{\beta-1} exp(-(x/\alpha)^\beta), \quad x > 0, \quad \alpha, \beta > 0$$

denoted by $X \sim \text{Weib}(\beta, \alpha)$. The cumulative distribution function (cdf) is,

$$F_X(x) = 1 - exp(-(x/\alpha)^\beta). \ x \geq 0$$

Now, let $X_1$, $X_2$, $X_3$ be three independent random variables following Weibull distributions with different parameters:

$$X_1 \sim \text{Weib}(1.2, 5), \quad X_2 \sim \text{Weib}(1.5, 6), \quad X_3 \sim \text{Weib}(2.6, 4)$$

with corresponding $pdfs$ $f_1(x)$, $f_2(x)$, $f_3(x)$ and cdfs $F_1(x)$, $F_2(x)$, $F_3(x)$. The generalized Weitzman overlapping coefficient between these three distributions is given by:

$$\Delta_3 = \int_0^\infty min\{f_1(x), f_2(x), f_3(x)\} dx.$$

This integral does not admit a closed-form expression in general due to the complexity introduced by the interaction of three different $pdfs$. As such, even in the case of only two distributions, analytical computation is often intractable without imposing restrictive assumptions on the parameters (e.g., equal shape parameters). In practical applications, numerical or simulation-based techniques are usually required.

To proceed, we approximate $\Delta_3$ numerically. First, we compute the approximate intersection points among the three $pdfs$ (see Figure 1):

- Let $a \cong 1.35$: intersection of $f_1(x)$ and $f_2(x)$.
- Let $b \cong 3.10$: intersection of $f_1(x)$ and $f_3(x)$.
- Let $c \cong 6.00$: intersection of $f_2(x)$ and $f_3(x)$

Based on these points, the integration domain $[0, \infty)$ is partitioned into four subintervals where the minimum function changes:

$$min\{f_1(x), f_2(x), f_3(x)\} = \begin{cases} f_3(x) &, \quad 0.00 \leq x < 1.35 \\ f_2(x) &, \quad 1.35 \leq x < 3.10 \\ f_1(x) &, \quad 3.10 \leq x < 6.00 \\ f_3(x) &, \quad 6.00 \leq x < \infty \end{cases}.$$

Hence, $\Delta_3$ can be evaluated numerically as:

$$\Delta_3 = \int_0^{1.35} f_3(x)dx + \int_{1.35}^{3.1} f_2(x)dx + \int_{3.1}^{6.0} f_1(x)dx + \int_{6.0}^\infty f_3(x)dx.$$

Each integral can be evaluated using the corresponding cdf:

$$\Delta_3 = F_3(1.35) - F_3(0) + F_2(3.1) - F_2(1.35) + F_1(6) - F_1(3.1) + F_3(\infty) - F_3(6)$$

$$= 1 - e^{-\left(\frac{1.35}{4}\right)^{2.6}} + e^{-\left(\frac{1.35}{6}\right)^{1.5}} - e^{-\left(\frac{3.1}{6}\right)^{1.5}} + e^{-\left(\frac{3.1}{5}\right)^{1.2}} - e^{-\left(\frac{6}{5}\right)^{1.2}} + e^{-\left(\frac{6}{4}\right)^{2.6}}$$

$$= 0.604503.$$

Thus, the overlapping coefficient $\Delta_3$ for these three Weibull distributions is approximately 0.6045. This example highlights both the utility and difficulty of calculating $\Delta_k$ directly, motivating the need for alternative estimation techniques.

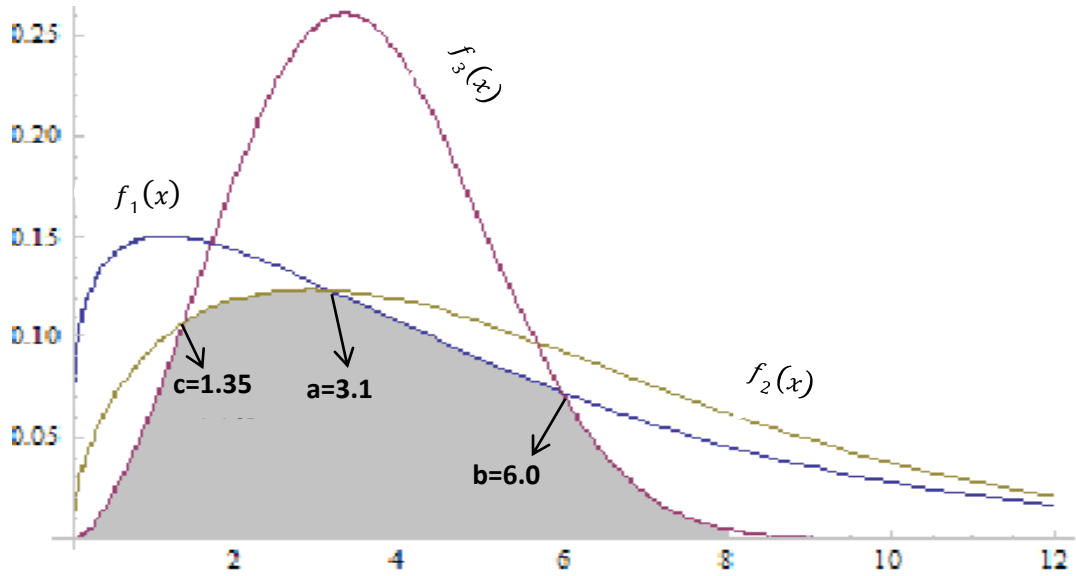

**Figure (1).** The Weitzman overlapping region $\Delta_3$ between the three Weibull distributions $f_1(x)$, $f_2(x)$ and $f_3(x)$.

### 3. Expressing the Weitzman Coefficient $\Delta_k$ as an Expectation

Let $X_1, X_2, \ldots X_k$ be independent continuous random variables with unknown $pdfs$ $f_1(x)$, $f_2(x), \ldots, f_k(x)$, respectively. Since the Weitzman overlapping coefficient $\Delta_k$ involves the integral of the pointwise minimum of these densities, it is often difficult to compute directly, particularly in nonparametric settings. To simplify the estimation, we can express $\Delta_k$ in terms of an expected value.

For any $i = 1, 2, \ldots, k$, define:

$$\Delta_k^{(i)} = E\left(\frac{min\{f_1(X_i), f_2(X_i), \ldots, f_k(X_i)\}}{f_i(X_i)}\right).$$

Since the expression inside the expectation is a function of $X_i$, we can write:

$$\Delta_k^{(i)} = \int_{-\infty}^{\infty} \frac{min\{f_1(x), f_2(x), \ldots, f_k(x)\}}{f_i(x)} f_i(x) dx = \Delta_k.$$

This implies,

$$\Delta_k^{(1)} = \Delta_k^{(2)} = \cdots = \Delta_k^{(k)} = \Delta_k$$

This shows that any single $\Delta_k^{(i)}$ yields an exact expression for $\Delta_k$, and all of them are equal. Therefore, a natural and practical expression is the average over all $k$ values:

$$\Delta_k = \frac{1}{k} \sum_{i=1}^{k} \Delta_k^{(i)} = \frac{1}{k} \sum_{i=1}^{k} E\left(\frac{min\{f_1(X_i), f_2(X_i), \ldots, f_k(X_i)\}}{f_i(X_i)}\right).$$

Furthermore, since all $\Delta_k^{(i)}$ are equals, $\Delta_k$ can also be expressed as the average of any subset of two, three, or more of these expressions. For example, when $k = 3$, we have,

$$\Delta_3^{(1,2)} = \frac{\Delta_3^{(1)} + \Delta_3^{(2)}}{2} = \Delta_3, \quad \Delta_3^{(1,3)} = \frac{\Delta_3^{(1)} + \Delta_3^{(3)}}{2} = \Delta_3, \quad \Delta_3^{(2,3)} = \frac{\Delta_3^{(2)} + \Delta_3^{(3)}}{2} = \Delta_3.$$

Similarly, averaging all three estimators yields:

$$\Delta_3^{(1,2,3)} = \frac{\Delta_3^{(1)} + \Delta_3^{(2)} + \Delta_3^{(3)}}{3}.$$

This flexibility offers multiple ways to estimate or approximate $\Delta_k$, which is especially useful in practice when dealing with sample-based estimators or computational constraints.

## 4. Estimating the Weitzman Coefficient $\Delta_k$

Let $X_1, X_2, \ldots X_k$ be independent continuous random variables with unknown $pdfs$ $f_1(x), f_2(x)$, $\ldots$, $f_k(x)$, respectively. That is, $X_i \sim f_i(x)$, $i = 1, 2, \ldots, k$. Suppose for each $i = 1, 2, \ldots, k$, we observe a random sample $X_{i1}, X_{i2}, \ldots, X_{in_i}$ of size $n_i$ from the distribution $f_i(x)$. These $k$ samples are assumed to be independent. Our objective is to estimate the Weitzman overlapping coefficient $\Delta_k$ using these samples.

### 4.1 Kernel Estimation of the Densities

To begin, we estimate each unknown density function $f_i(x)$ using the kernel density estimator. If the support of the data is $(-\infty, \infty)$, the standard kernel estimator at point $x$ is:

$$\hat{f}_i(x) = \frac{1}{n_i h_i} \sum_{m=1}^{n_i} K\left(\frac{x - X_{im}}{h_i}\right), \quad i = 1, 2, \ldots, k,$$

where $K(.)$ is a kernel function and $h_i$ is the bandwidth parameter. For data supported on $(-\infty, \infty)$, we use the reflection kernel method (Silverman, 1986):

$$\hat{f}_i(x) = \frac{1}{n_i h_i} \sum_{m=1}^{n_i} \left\{ K\left(\frac{x - X_{im}}{h_i}\right) + K\left(\frac{x + X_{im}}{h_i}\right) \right\}.$$

### 4.2 Estimating $\Delta_k$ Using the Method of Moments

Recall from the previous section that $\Delta_k$ can be expressed as:

$$\Delta_k = \Delta_k^{(i)} = E\left(\frac{min\{f_1(X_i), f_2(X_i), \ldots, f_k(X_i)\}}{f_i(X_i)}\right), \quad i = 1, 2, \ldots, k.$$

Using the method of moments, we approximate this expectation by the sample average:

$$\hat{\Delta}_k^{(i)} = \frac{1}{n_i} \sum_{j=1}^{n_i} \frac{min\{\hat{f}_1(X_{ij}), \hat{f}_2(X_{ij}), \ldots, \hat{f}_k(X_{ij})\}}{\hat{f}_i(X_{ij})}, \quad i = 1, 2, \ldots, k$$

Each $\hat{\Delta}_k^{(i)}$ is a consistent estimator of the true overlapping coefficient $\Delta_k$. Since all theoretical values $\Delta_k^{(i)}$ are equal, it is also reasonable to consider averaging two or more the individual estimators to construct alternative estimators of $\Delta_k$ with potential improved performance.

In the case where $k = 3$, we obtain the following seven nonparametric estimators for $\Delta_3$,

$$\hat{\Delta}_3^{(1)}, \hat{\Delta}_3^{(2)}, \hat{\Delta}_3^{(3)}, \hat{\Delta}_3^{(1,2)}, \hat{\Delta}_3^{(1,3)}, \hat{\Delta}_3^{(2,3)}, \hat{\Delta}_3^{(1,2,3)}$$

These estimators correspond to the parameters $\Delta_3^{(1)}, \Delta_3^{(2)}, \Delta_3^{(3)}, \Delta_3^{(1,2)}, \Delta_3^{(1,3)}, \Delta_3^{(2,3)}$ and $\Delta_3^{(1,2,3)}$, respectively.

### 4.3 Averaged Estimator of $\Delta_k$

A commonly used and practical estimator is the average of all $k$ estimators $\hat{\Delta}_k^{(i)}$, defined as,

$$\Delta_{avg} = \frac{1}{k} \sum_{i=1}^{k} \hat{\Delta}_k^{(i)} = \frac{1}{k} \sum_{i=1}^{k} \left( \frac{1}{n_i} \sum_{j=1}^{n_i} \frac{min\{\hat{f}_1(X_{ij}), \hat{f}_2(X_{ij}), \ldots, \hat{f}_k(X_{ij})\}}{\hat{f}_i(X_{ij})} \right).$$

This estimator balances contributions from all distributions and is particularly useful when the sample sizes $n_i$ vary.

This estimation framework forms the basis for developing consistent and nonparametric estimators of the Weitzman coefficient $\Delta_k$, using observed samples and kernel smoothing techniques.

## 5. Monte Carlo Simulation Study

In this section, we design and conduct a simulation study to evaluate the performance of the proposed nonparametric estimators of the Weitzman overlapping coefficient, $\Delta_k$. Specifically, we consider the case $k = 3$, which requires generating three random samples from three distinct distributions to compute and assess a subset of the seven estimators we developed: $\Delta_3^{(1)}$, $\Delta_3^{(2)}$, $\Delta_3^{(3)}$, $\Delta_3^{(1,2)}$, $\Delta_3^{(1,3)}$, $\Delta_3^{(2,3)}$ and $\Delta_3^{(1,2,3)}$ (see Subsection 4.2). For brevity we focus our analysis on four representative estimators: $\Delta_3^{(1)}$, $\Delta_3^{(1,2)}$, $\Delta_3^{(2,3)}$ and $\Delta_3^{(1,2,3)}$. Preliminary simulation results (not shown) indicate that the remaining three estimators yield similar conclusions.

Since the proposed estimators are nonparametric, we generate samples from three different parametric families of distributions to assess their robustness. One distribution is supported on the positive real line, while the other two are supported on the entire real line. All simulations were conducted using *Mathematica, Version 7*, for the following combinations of sample sizes: $(n_1, n_2, n_3) = (50, 50, 50), (50, 100, 150)$. The selected distribution families are:

- Normal distribution, $N(\mu, \sigma^2)$:

$$f(x) = \frac{1}{\sqrt{2\pi\sigma^2}} e^{\frac{(x-\mu)^2}{2\sigma^2}}, \ x \in \mathcal{R}, \ \mu \in \mathcal{R}, \ \sigma^2 > 0$$

- Extreme value Distribution, $Ex(\mu, \alpha)$:

$$f(x) = \frac{1}{\alpha} e^{(\frac{x-\mu}{\alpha} - e^{\frac{x-\mu}{\alpha}})}, \ x \in \mathcal{R}, \ \alpha > 0,$$

- Weibull distribution, $W(\beta, \alpha)$:

$$f(x) = \frac{\beta}{\alpha} \left(\frac{x}{\alpha}\right)^{\beta-1} e^{-(x/\alpha)^\beta}, \quad x > 0, \quad \alpha, \beta > 0$$

For each distribution family, four different parameter settings were chosen to produce a range of exact $\Delta_3$ values, from high to low. Details of each setting, including the corresponding exact values of $\Delta_3$ are summarized in Table (1). The Gaussian kernel was used as the kernel function $K$ for all nonparametric estimators. The smoothing parameter $h_i$ for each sample was selected using Silverman's rule of thumb (1986):

$$h_i = (4/3)^{1/5} A_i n_i^{-1/5}, \quad i = 1, 2, 3$$

where

$$A_i = min\left\{\text{standard deviation}, \frac{\text{interquartile range}}{1.34}\right\}$$

(See Section 1). Each bandwidth $h_i$ is computed separately for each sample, based on its distribution: $f_1(x)$, $f_2(x)$ and $f_3(x)$, respectively.

To evaluate the performance of each proposed estimator, we computed the following metrics over 1000 simulation replicates (see also, Eidous and Alshyyab, 2025, 2026)**:**

- The relative bias ($RB$): $RB = (\hat{E}(estimator) - exact)/exact$.
- The relative root of mean squared error ($RMSE$): $RMSE = \sqrt{\widehat{MSE}(estimator)}/exact$.

**Table (1)**. Simulated distributions and exact values of $\Delta_3$.

| Distribution | $f_1(x)$ | $f_2(x)$ | $f_3(x)$ | Exact $\Delta_3$ |
|---|---|---|---|---|
| Case 1: Normal | $N(-0.1,1)$ | $N(0,0.9)$ | $N(0.2,1.1)$ | 0.8602 |
| Case 2: Normal | $N(-0.4,1)$ | $N(0.1,0.88)$ | $N(0.4,1.4)$ | 0.6550 |
| Case 3: Normal | $N(-0.5,1)$ | $N(0,0.5)$ | $N(0.75,1)$ | 0.4688 |
| Case 4: Normal | $N(-1,1.5)$ | $N(0,0.8)$ | $N(2,0.4)$ | 0.0735 |
| Case 5: Extreme | $Ex(0,1)$ | $Ex(1.25,1.2)$ | $Ex(1,1.2)$ | 0.9159 |
| Case 6: Extreme | $Ex(0,1)$ | $Ex(1.5,2)$ | $Ex(1,1.4)$ | 0.6397 |
| Case 7: Extreme | $Ex(0,1)$ | $Ex(1.5,3)$ | $Ex(1,1)$ | 0.4227 |
| Case 8: Extreme | $Ex(0.5,0.2)$ | $Ex(2,6)$ | $Ex(2,0.75)$ | 0.0878 |
| Case 9: Weibull | $W(1.1,1)$ | $W(1.25,1.2)$ | $W(1,1.20)$ | 0.8626 |
| Case 10: Weibull | $W(1.5,1)$ | $W(1.5,2)$ | $W(1,1.4)$ | 0.5984 |
| Case 10: Weibull | $W(1.5,1)$ | $W(1.5,3)$ | $W(1,1)$ | 0.4429 |
| Case 11: Weibull | $W(2,1)$ | $W(2,6)$ | $W(2,0.75)$ | 0.0785 |

## 6. Simulation Results

The simulation results for the proposed nonparametric estimators of the overlapping coefficient $\Delta_3$ are summarized in Tables 2–4. These tables present the average estimate, relative bias (RB), and relative root mean square error (RMSE) across various distributional cases (Normal, Extreme, and Weibull) and sample sizes. Based on a comprehensive examination of these results, we highlight the following key findings:

**I. Bias Trends and Potential for Improvement**

Across most scenarios, the relative bias (RB) values of the proposed estimators are negative, indicating a general tendency to underestimate the true value of $\Delta_3$. This underestimation is consistent regardless of the underlying distribution (normal, extreme, or Weibull). The persistent negative bias suggests the presence of inherent bias in the traditional kernel method used for estimation. As a result, future work may benefit from exploring bias-reduction techniques, such as fourth-order kernel estimators (Wand and Jones, 1995) or adaptive (variable) kernel methods (Breiman et al., 1977), both of which have been shown to effectively address bias in nonparametric density estimation.

## II. Consistency of the Estimators

A consistent trend across all cases is the decrease in RMSE values with increasing sample sizes. This result is expected and confirms that the proposed estimators satisfy the important statistical property of consistency, i.e., their accuracy improves as the sample size grows. For example, in Case 1 (Normal, $\Delta_3$=0.8602), RMSE decreases from 0.1442 at small sample size, to 0.1036 at moderate sample size for estimator $\hat{\Delta}_3^{(1)}$.

## III. Performance Comparison and Efficiency

The relative performance of the different estimators is generally similar, as evidenced by closely aligned RMSE values. However, some fluctuations in RMSE are observed across certain cases:

Estimators such as $\hat{\Delta}_3^{(1)}$ and $\hat{\Delta}_3^{(1,2)}$ or $\hat{\Delta}_3^{(2,3)}$ outperform the overall estimator $\hat{\Delta}_3^{(1,2,3)}$ in specific settings. For example, in Case 8 (small sample size), the RMSE value for $\hat{\Delta}_3^{(1)}$ is less than that of $\hat{\Delta}_3^{(1,2,3)}$, indicating superior performance compared to the baseline.

Conversely, in other cases (e.g., Cases 9 and 11), the same estimator exhibits RMSE values significantly geater than that of $\hat{\Delta}_3^{(1,2,3)}$, reflecting relatively poorer performance.

These variations underscore the importance of considering case-specific behavior when selecting an estimator. Nevertheless, the estimator $\hat{\Delta}_3^{(1,2,3)}$ consistently demonstrates stable performance across all scenarios, with relatively moderate bias and RMSE in most cases. This stability supports its use as a robust, general-purpose estimator of $\Delta_3$.

## IV. Dependence of RMSE on the Magnitude of $\Delta_3$

A notable observation is the inverse relationship between the magnitude of the true overlapping coefficient $\Delta_3$ and the RMSE of its estimators. Specifically, smaller values of $\Delta_3$ are associated with higher RMSE values. For example, in Case 4 (Normal, $\Delta_3$=0.0735), RMSE values are as high as 0.4177 for $\hat{\Delta}_3^{(1,2,3)}$, whereas for larger $\Delta_3$ values (e.g., Case 5, Extreme, $\Delta_3$=0.9159), the RMSE is substantially lower (0.1592 for the same estimator with the same sample size). This pattern suggests that the proposed estimators are more reliable when estimating moderate to large overlap coefficients, but their performance deteriorates in the presence of minimal overlap.

## V. Recommendation

While all the proposed estimators show satisfactory performance, we recommend using the estimator $\hat{\Delta}_3^{(1,2,3)}$ as the primary estimator for $\Delta_3$. This recommendation is based on its consistent performance, relatively low RMSE across various distributional settings, and stable behavior across all simulated cases. It offers a balanced trade-off between bias and variance and is particularly useful when no prior knowledge about the distributions is available.

**Table (2).** Average, relative bias (RB) and relative root of mean square error (RMSE) of the various estimators of $\Delta_3$ based on data simulated from three **normal** distributions.

| $(n_1, n_2, n_3)$ | | $\hat{\Delta}_3^{(1)}$ | $\hat{\Delta}_3^{(1,2)}$ | $\hat{\Delta}_3^{(2,3)}$ | $\hat{\Delta}_3^{(1,2,3)}$ |
|---|---|---|---|---|---|
| | | **Case 1: $\Delta_3$ = 0.8602** | | | |
| (50,50,50) | Average | 0.7517 | 0.7626 | 0.7573 | 0.7572 |
| | RB | -0.1260 | -0.1140 | -0.1200 | -0.1200 |
| | RMSE | 0.1442 | 0.1315 | 0.1376 | 0.1376 |
| (50,100,150) | Average | 0.786 | 0.7948 | 0.7907 | 0.7905 |
| | RB | -0.0863 | -0.076 | -0.0808 | -0.0810 |
| | RMSE | 0.1036 | 0.0944 | 0.0981 | 0.0986 |
| | | **Case 2: $\Delta_3$ = 0.6550** | | | |
| (50,50,50) | Average | 0.6184 | 0.6359 | 0.6296 | 0.628 |
| | RB | -0.0558 | -0.0291 | -0.0387 | -0.0412 |
| | RMSE | 0.1230 | 0.1142 | 0.1156 | 0.1169 |
| (50,100,150) | Average | 0.6343 | 0.6469 | 0.6439 | 0.6417 |
| | RB | -0.0316 | -0.0124 | -0.017 | -0.0203 |
| | RMSE | 0.0955 | 0.0920 | 0.0899 | 0.0920 |
| | | **Case 3: $\Delta_3$ = 0.4688** | | | |
| (50,50,50) | Average | 0.4458 | 0.4442 | 0.4274 | 0.4335 |
| | RB | -0.0490 | -0.0525 | -0.0883 | -0.0752 |
| | RMSE | 0.1508 | 0.1505 | 0.1606 | 0.1551 |
| (50,100,150) | Average | 0.4572 | 0.4546 | 0.4427 | 0.4475 |
| | RB | -0.0247 | -0.0302 | -0.0556 | -0.0453 |
| | RMSE | 0.1152 | 0.1095 | 0.1122 | 0.1111 |
| | | **Case 4: $\Delta_3$ = 0.0735** | | | |
| (50,50,50) | Average | 0.0582 | 0.0582 | 0.0571 | 0.0574 |
| | RB | -0.2080 | -0.2080 | -0.2230 | -0.2180 |
| | RMSE | 0.4924 | 0.4267 | 0.4046 | 0.4177 |
| (50,100,150) | Average | 0.0616 | 0.0646 | 0.0654 | 0.0641 |
| | RB | -0.1620 | -0.1200 | -0.1100 | -0.1270 |
| | RMSE | 0.4267 | 0.3358 | 0.2898 | 0.3184 |

**Table (3).** Average, relative bias (RB) and relative root of mean square error (RMSE) of the various estimators of $\Delta_3$ based on data simulated from three **extreme** distributions.

| $(n_1, n_2, n_3)$ | | $\hat{\Delta}_3^{(1)}$ | $\hat{\Delta}_3^{(1,2)}$ | $\hat{\Delta}_3^{(2,3)}$ | $\hat{\Delta}_3^{(1,2,3)}$ |
|---|---|---|---|---|---|
| | | **Case 5: $\Delta_3$ = 0.9159** | | | |
| (50,50,50) | Average | 0.7705 | 0.7820 | 0.7875 | 0.7818 |
| | RB | -0.1590 | -0.1460 | -0.1400 | -0.146 |
| | RMSE | 0.1725 | 0.1590 | 0.1531 | 0.1592 |
| (50,100,150) | Average | 0.8126 | 0.8234 | 0.8295 | 0.8239 |
| | RB | -0.1130 | -0.1010 | -0.0943 | -0.101 |
| | RMSE | 0.1268 | 0.1135 | 0.1063 | 0.1128 |
| | | **Case 6: $\Delta_3$ = 0.6397** | | | |
| (50,50,50) | Average | 0.5861 | 0.6114 | 0.6225 | 0.6104 |
| | RB | -0.0839 | -0.0444 | -0.0269 | -0.0459 |
| | RMSE | 0.1435 | 0.1221 | 0.1163 | 0.1225 |
| (50,100,150) | Average | 0.6001 | 0.6233 | 0.6347 | 0.6232 |
| | RB | -0.062 | -0.0257 | -0.0078 | -0.0259 |
| | RMSE | 0.1191 | 0.1028 | 0.0970 | 0.1013 |
| | | **Case 7: $\Delta_3$ = 0.4227** | | | |
| (50,50,50) | Average | 0.3903 | 0.4128 | 0.4055 | 0.4005 |
| | RB | -0.0766 | -0.0234 | -0.0406 | -0.0526 |
| | RMSE | 0.1568 | 0.1462 | 0.1471 | 0.1485 |
| (50,100,150) | Average | 0.3994 | 0.4187 | 0.4155 | 0.4101 |
| | RB | -0.0551 | -0.0093 | -0.0170 | -0.0297 |
| | RMSE | 0.1329 | 0.1253 | 0.1222 | 0.1240 |
| | | **Case 8: $\Delta_3$ = 0.0878** | | | |
| (50,50,50) | Average | 0.0653 | 0.0691 | 0.0708 | 0.0690 |
| | RB | -0.2560 | -0.2130 | -0.1930 | -0.2140 |
| | RMSE | 0.3479 | 0.364 | 0.3819 | 0.3633 |
| (50,100,150) | Average | 0.0656 | 0.0704 | 0.0731 | 0.0707 |
| | RB | -0.2520 | -0.197 | -0.166 | -0.195 |
| | RMSE | 0.3104 | 0.2945 | 0.2905 | 0.2910 |

**Table (4).** Average, relative bias (RB) and relative root of mean square error (RMSE) of the various estimators of $\Delta_3$ based on data simulated from three **Weibull** distributions.

| $(n_1, n_2, n_3)$ | | $\hat{\Delta}_3^{(1)}$ | $\hat{\Delta}_3^{(1,2)}$ | $\hat{\Delta}_3^{(2,3)}$ | $\hat{\Delta}_3^{(1,2,3)}$ |
|---|---|---|---|---|---|
| | | **Case 9: $\Delta_3$ = 0.8626** | | | |
| (50,50,50) | Average | 0.7682 | 0.7712 | 0.7746 | 0.7724 |
| | RB | -0.1100 | -0.1060 | -0.1020 | -0.1050 |
| | RMSE | 0.1296 | 0.1254 | 0.1217 | 0.1241 |
| (50,100,150) | Average | 0.7949 | 0.8003 | 0.8063 | 0.8025 |
| | RB | -0.0787 | -0.0724 | -0.0656 | -0.0700 |
| | RMSE | 0.1002 | 0.0936 | 0.0873 | 0.0914 |
| | | **Case 10: $\Delta_3$ = 0.5984** | | | |
| (50,50,50) | Average | 0.5553 | 0.5687 | 0.5736 | 0.5675 |
| | RB | -0.0720 | -0.0496 | -0.0415 | -0.0500 |
| | RMSE | 0.1343 | 0.1243 | 0.1235 | 0.1259 |
| (50,100,150) | Average | 0.5626 | 0.5759 | 0.5824 | 0.5758 |
| | RB | -0.0598 | -0.0376 | -0.0267 | -0.0377 |
| | RMSE | 0.1097 | 0.0991 | 0.0961 | 0.0992 |
| | | **Case 11: $\Delta_3$ = 0.4429** | | | |
| (50,50,50) | Average | 0.3971 | 0.4094 | 0.4110 | 0.4064 |
| | RB | -0.1030 | -0.0756 | -0.0720 | -0.0824 |
| | RMSE | 0.1703 | 0.1616 | 0.16040 | 0.1625 |
| (50,100,150) | Average | 0.4031 | 0.4157 | 0.4205 | 0.4147 |
| | RB | -0.0899 | -0.0613 | -0.0504 | -0.0636 |
| | RMSE | 0.1420 | 0.1293 | 0.1242 | 0.1286 |
| | | **Case 12: $\Delta_3$ = 0.0785** | | | |
| (50,50,50) | Average | 0.0842 | 0.0756 | 0.0726 | 0.0765 |
| | RB | 0.0725 | -0.0367 | -0.0747 | -0.0256 |
| | RMSE | 0.3604 | 0.3941 | 0.3941 | 0.3697 |
| (50,100,150) | Average | 0.0846 | 0.0785 | 0.0794 | 0.0812 |
| | RB | 0.0779 | 0.0001 | 0.0115 | 0.0336 |
| | RMSE | 0.2738 | 0.2803 | 0.2837 | 0.2733 |

**Declaration:** The authors state that there is no conflict of interest. We used ChatGPT to refine and correct the language.